\documentstyle[12pt,epsfig]{article}

\begin{document}
\title{Comment on New Results for the Charged Pion
Electromagnetic Form Factor}
\author{ A.~F.~Krutov\thanks{E-mail:
krutov@info.ssu.samara.ru}\\
{\small {\em Samara State
University, 443011, Samara, Russia }}\\
V.~E.~Troitsky\thanks{E-mail:
troitsky@theory.npi.msu.su}\\
{\small {\em Nuclear Physics
Institute, Moscow State University, 119899, Moscow, Russia }} }
\date{October 23, 2000}
\maketitle
\begin{abstract}
In the paper nucl-ex/0010009  by The Jefferson   Lab $F_\pi$
Collaboration the new results for the charged pion
electromagnetic form factor are presented.  The approach given
in our paper hep-ph/9811318, "On a possible estimation of the
constituent--quark parameters from Jefferson Lab experiments on
the pion form factor" makes it possible to estimate the mass of
constituent quark from such kind of experiment. Now this
estimation gives $M = 0.21$ GeV.
\end{abstract}

In the paper ~\cite{1}  by The Jefferson   Lab $F_\pi$
Collaboration the new results for the charged pion
electromagnetic form factor are presented.  The form factor was
measured in momentum transfer region $Q^2 = 0.6 - 1.6 $
(GeV/c)$^2$ .  The results indicate that the pion form factor in
this region is larger than previously assumed.

Some time ago using one of relativistic forms of the
constituent quark model (CQM) we  presented a method of
calculation of the pion electromagnetic form factor ~\cite{2}.
The results of our calculation indicate that the form factor
dependence on the quark interaction model is weak, while the
dependence on the constituent quark mass is rather significant.
The curves calculated with different wave functions but one and
the same quark mass form groups. The position of the group
changes essentially with the quark mass.  So the approach [2]
makes it possible to estimate the mass of constituent quark from
the experiment on pion form factor.

Let us remind that we have fixed the parameters of the model in
the following way.  At the present time there are two pion
characteristics that can be extracted from the data in a model
independent way and with sufficient accuracy:  the mean square
radius $ \langle r_\pi^2\rangle ^{1/2}_{exp}$, and the lepton
decay constant $ f_{\pi\>exp}$.  We assume that the calculations
for any quark interaction model satisfy (in addition to the
description of the particle spectrum) the conditions:
$$
\langle r_\pi^2\rangle ^{1/2} = \langle r_\pi^2\rangle
^{1/2}_{exp},
$$
$$ f_\pi = f_{\pi\>exp}.
$$
These characteristics are not influenced by the new data.

\begin{figure*}
\begin{picture}(0,0)(0,250)%
\centerline{\epsfxsize=0.4\textwidth \epsfbox{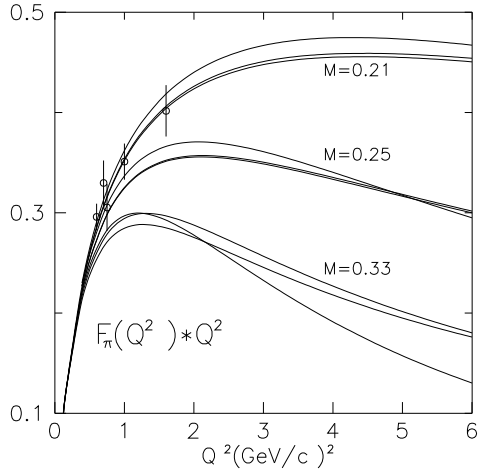}}
\end{picture}%
\vspace{8.5cm}
\caption{$\pi$--meson form factor in the range of Jefferson Lab
experiments. The results of calculations for different
interaction models and $M$=0.21, 0.25, 0.33 GeV.
The curves with the same mass form a group.
The position of a group is defined by the constituent quark
mass.}
\end{figure*}
Our results are presented in Fig.1. Instead of the group
with $M = 0.22$ GeV in ~\cite{2} we present the group with
$M = 0.21$ GeV providing the best description
of the new data.

The great accuracy of Jefferson Lab experiments ~\cite{1} makes
it possible to fix the position of "the group" rather accurately
and, so, to determine the constituent quark mass $M$.

So, the result is: the constituent quark mass $M = 0.21$ GeV.

\end{document}